# Article for Nature Astronomy - News & Views

GAMMA-RAY BURSTS

# To be short or long is not the question

The association of a short gamma-ray burst with a core-collapse supernova seems to challenge current scenarios for the origin of these extreme events. But how much can we rely on observed duration for pinpointing their progenitors?

## Lorenzo Amati

Gamma-ray bursts (GRBs) are extremely bright flashes of X-/ gamma-ray photons detected about once per day from random directions and outshining for up to a few minutes any other source in the high-energy sky. Despite huge observational efforts, it took as long as thirty years from their discovery in the 1960s to unveil their cosmological origin. A further twenty years of efforts by many space and ground observatories, as well as intensive theoretical work and numerical simulations, were then needed to build up and consolidate the current scenario for their progenitors: core collapse of peculiar very massive stars for the longer ones and merging of a neutron star–neutron star (NS–NS) or neutron star–black hole (NS–BH) binary system for the shorter ones[1,2]. Writing in *Nature Astronomy*, Tomás Ahumada[3], Bin-Bin Zhang[4] and their collaborators report significant evidence for the unexpected association of a short GRB with a stellar explosion. Do we actually need to reconsider our main paradigm for GRBs?

The mounting evidence for the existence of short (a few tens of milliseconds up to 1–2 seconds) and long (tens to hundreds of seconds) GRBs was one of the first main steps forward in our comprehension of these outstanding but elusive phenomena, based on the distribution of their durations (Fig. 1d) and on the spectral hardness– duration plane (Fig. 1a)[5,6]. A major step forward occurred towards the dawn of the new millennium, when the first systematic arcmin localizations of long bursts allowed the discovery of the fading multi-wavelength GRB 'afterglow' emission[8,9]. The improved (a few arcsec) localization and the spectroscopy of the optical/near infrared afterglows by large ground telescopes and the Hubble Space Telescope[7,8] led to establishing the cosmological distance scale of long GRBs (extending up to a redshift of at least ~9), the first identifications and characterizations of their host galaxies[9,10] and the direct detection of a peculiar type Ib/c supernova (SN) associated with the long GRB 980425[11]. This impressive wealth of discoveries provided strong support for the production of long GRBs by the core collapse of peculiar massive stars, a scenario already postulated in the 1970s. Such a core collapse is capable of explaining the long duration and extremely high radiated energy (up to $10^{53}$erg or even more)[12] and was further strengthened in the last 20 years by the detection of a 'bump' over the decaying afterglow light curves with spectral features resembling those of peculiar type Ib/c core-collapse supernovas[1]. Other evidence supporting this origin for long GRBs includes their typical location in active star-forming regions of their host galaxies (Fig. 1e), the evidence of a metal-enriched circum-burst environment, and their redshift distribution approximating the star-formation rate evolution.

Conversely, from the early 2000s, we started learning that for short GRBs the redshift distribution extends to much lower redshift than long ones, that their typical released energy is about two orders of magnitude lower and that they are often located in the outskirts of their host galaxies (Fig. 1c) without any association with star-forming regions[2]. Together with their duration, these properties support the hypothesis that short GRBs originate from the merging of NS–NS or NS–BH binary systems[1,2], a scenario that was eventually confirmed by the association of a gravitational wave signal from a NS–NS merger (GW170817) and a short GRB (170817A)[13].

The observations and analyses reported by Ahumada et al. and Zhang et al., however, seem to defy expectations: a SN bump was detected in the afterglow light curve of the genuinely short GRB 200826A. Although at face value this result may challenge our current understanding of the GRB phenomenon, a more global view of its properties, combined with several pieces of evidence that have emerged in the last few years, show that this event, while peculiar and of high interest, may not be so odd.

First of all, the ~1 s duration of GRB 200826A is in the range where the duration distributions of short and long GRBs still overlap (Fig. 1d), potentially making this an extreme event in the short-duration tail of the long GRB duration distribution. The location in the spectral hardness– duration plane further supports a short GRB classification, but the probability that GRB 200826A belongs to the long class is still not negligible[2]. Moreover, the divide between the two distributions varies as a function of the energy band considered and, more generally, the characteristics of the GRB detector used to build the sample from which the distributions are drawn. To accentuate the problem, theoretical considerations and numerical simulations show that the duration of a core-collapse GRB, which depends on the time during which the 'central engine' (the BH accreting matter from the torus) is at work and the time taken by the produced jet to break out from the stellar envelope, can be even shorter than 0.5 s (ref. [14]).

This is why duration is increasingly considered to be only one of the indicators of the origin of a GRB. Instead, these two classes of events are now frequently referred to as 'type I' and 'type II'[3,4]. In addition to duration, spectral hardness, location in the host galaxy and host-galaxy properties, the indicators used for discriminating the two classes of events include the 'time lag', the delay of the peak of the emission as a function of the energy band, and the relation between the photon energy at which the $\nu F_\nu$ spectrum in the cosmological rest-frame peaks ($E_{p,i}$) and the isotropic-equivalent radiated energy ($E_{iso}$) . The analysis of these indicators for GRB 200826A shows that despite its short duration, it actually belongs to the type II/ core -collapse class. Thus, the association of this event with a type Ib/c SN is no longer surprising, and instead provides a strong confirmation of the efficiency of our new paradigm for identifying the progenitor of a GRB. In this respect, GRB 200826A is the opposite case of another famous and challenging event, GRB 060614, which was technically a long GRB for which there was

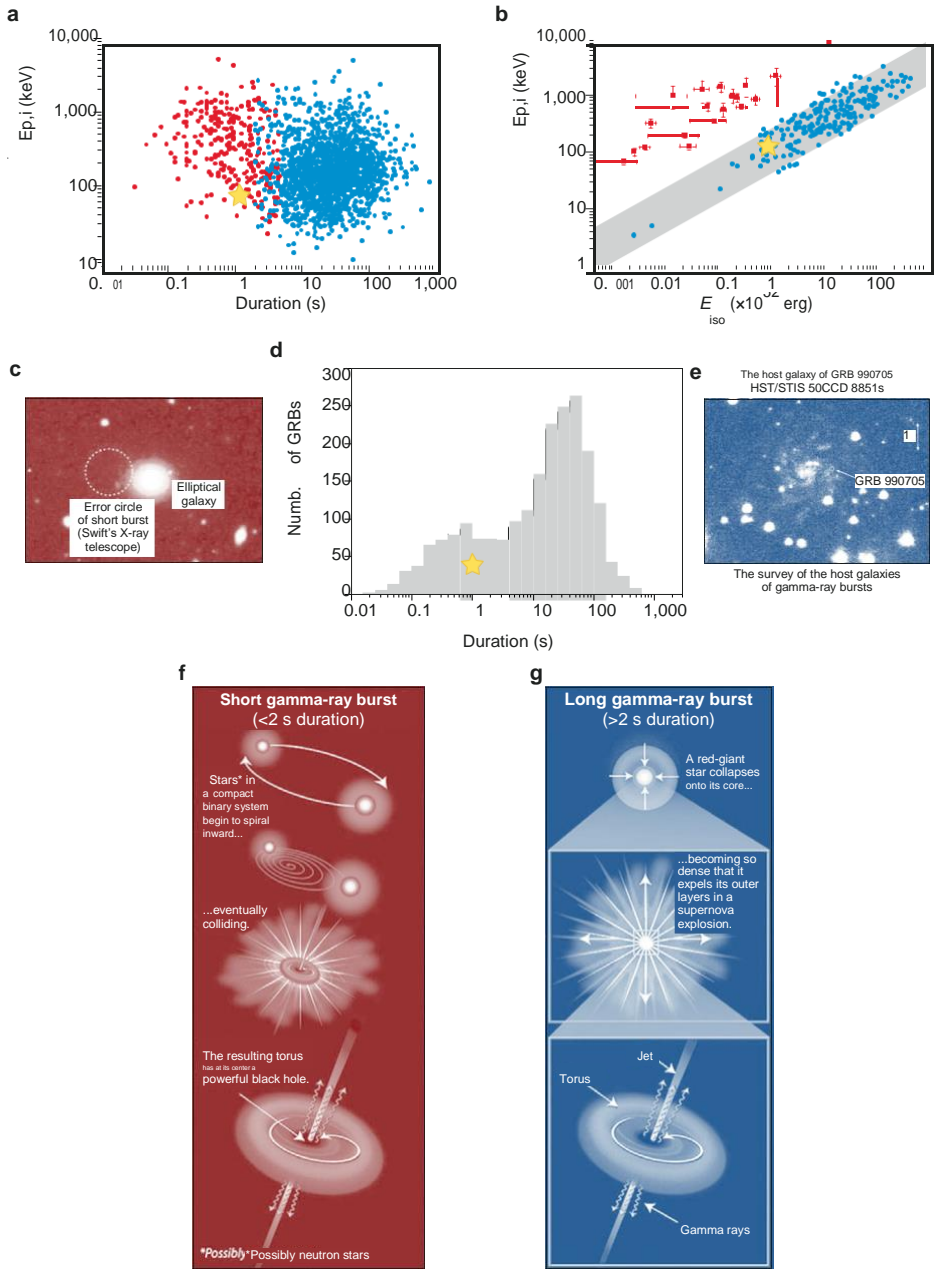

The extended sensitivity and energy bands of next-generation GRB detectors (for example, those on board the Chinese–French mission, the Space Variable Objects Monitor, to be launched next year), combined with improved follow-up capabilities provided by future ground observatories (for example, the Extremely Large Telescope or the Thirty Meter Telescope in the optical/ near infrared, and the Square Kilometre Array in the radio), will allow us to eventually assess whether currently peculiar events like the odd balls GRB 200826A and GRB 060614, the fraction of short GRBs showing a soft extended emission, under-luminous low-redshift long GRBs (for example, GRB 980425) and the few ultra-long (thousands of seconds) GRBs are actually tip-of-the-iceberg phenomena from large populations still to be unveiled. This substantial progress will allow us to definitively go beyond the GRB classification based on duration and probably to pinpoint a richer variety of progenitors, including extremely magnetized neutron stars for short GRBs as well as different types of progenitor stars (possibly including population III ones) and explosive processes (for example, pair instability) for long ones. An improved understanding of GRB subclasses and progenitors is also crucial for the growing relevance of type II events for cosmology (investigating the early Universe and possibly cosmological parameters) and of type I events for multi-messenger astrophysics

(as demonstrated by the amazing case of GW 170817/GRB 170817A).

Lorenzo Amati

*Italian National Institute for Astrophysics (INAF – OAS Bologna), Bologna, Italy.*

E-mail:lorenzo.amati@inaf.it

**Fig. 1 | Main observational differences between short/type I and long/type II GRBs and standard scenarios for their origin. a–g,** The bimodal distribution of the durations of GRBs (**d**), while providing evidence for the existence of two classes of GRBs, is not as efficient as may be expected in identifying the progenitor of a given event as a NS–NS (or NS–BH) merger (type I; **f**) or a core-collapsing star (type II; **g**). For this very important classification task, further observational features need to be considered, like the GRB location within its host galaxy (**c**,**e**), the properties of the host galaxy itself, and, importantly, the main characteristics of the prompt X-/gamma-ray emission, including the location in the spectral hardness–duration (**a**) and $E_{p,i}$–$E_{iso}$ (**b**) planes. Panel **a** is based on data from the Fermi/GBM online catalogue, **b** on data from BeppoSAX, HETE-2, Swift/BAT, Konus-WIND and Fermi/GBM, and **d** on data from the CGRO/BATSE 4B catalogue. Credit: Cristiano Guidorzi (University of Ferrara) (**a**,**b**,**d**); Joshua Bloom, C. Blake, J. X. Prochaska, J. Hennawi, M. Gladders and B. Koester (**c**); HST/STIS and Stephen Holland (**e**); NASA and A. Feild (STScI) (**f**,**g**).

strong evidence of no association with a SN[15]. In that case, similarly, it was thanks indicators other than duration (e.g., time lag and location in the $E_{p,i}$–$E_{iso}$ plane) that it was possible to solve the mystery by classifying the event as a type I burst.